\documentclass[conference]{IEEEtran}
\IEEEoverridecommandlockouts
\usepackage{cite}
\usepackage{amsmath,amssymb,amsfonts}
\usepackage{algorithmic}
\usepackage{graphicx}
\usepackage{textcomp}
\usepackage{xcolor}
\def\BibTeX{{\rm B\kern-.05em{\sc i\kern-.025em b}\kern-.08em
    T\kern-.1667em\lower.7ex\hbox{E}\kern-.125emX}}
\begin{document}

\title{Security of Virtual Reality Authentication Methods in Metaverse: An Overview
}

\author{\IEEEauthorblockN{Pınar Kürtünlüoğlu}
\IEEEauthorblockA{\textit{}
Computer Engineering\\
Mugla Sitki Kocman University\\
Mugla,Turkiye \\
pinarkurtunluoglu@posta.mu.edu.tr}
https://orcid.org/0000-0003-4525-1957
\and
\IEEEauthorblockN{Beste Akdik }
\IEEEauthorblockA{\textit{}
Computer Engineering\\
Mugla Sitki Kocman University\\
Mugla,Turkiye \\
besteakdik@posta.mu.edu.tr}
https://orcid.org/0000-0002-6691-445X

\and
\IEEEauthorblockN{Enis Karaarslan}
\IEEEauthorblockA{\textit{Department of Computer Engineering} \\
\textit{Muğla Sıtkı Koçman University}\\
Muğla, Turkey\\
Email: enis.karaarslan@mu.edu.tr\\
https://orcid.org/0000-0002-3595-8783}

}
\maketitle

\begin{abstract}
The metaverse is said to be the future Internet and will consist of several worlds called verses. This concept is being discussed a lot lately, however, the security issues of these virtual worlds are not discussed enough. This study first discusses the privacy and security concerns of the metaverse. Virtual reality headsets are the main devices used to access the Metaverse. The user needs to verify their identity to log in to the metaverse platforms, and the security of this phase becomes vital. This paper aims to compare the security of the main authentication methods that are used in virtual reality environments. Information-based, biometric, and multi-model methods are compared and analyzed in terms of security. These methods aim to verify the user with different data types such as 3D patterns, PIN systems, or biometric data. The pros and cons are discussed. The paper also concludes with what work can be done to improve the safety of these authentication methods and future work.
\end{abstract}

\begin{IEEEkeywords}
  virtual reality, metaverse, authentication, security, privacy 
\end{IEEEkeywords}

\section{INTRODUCTION}
The term "Metaverse" was actually used for the first time in 1992, used by Neal Stephenson, a student at Boston University who is very interested in computers, in his 1992 novel Snow Crash, even though we define it exactly on this day. In the novel, avatars and VR glasses are also mentioned, as in the present day. The metaverse promises more experiences than just the time we spend on regular social networking platforms. There are promising projects which are called pre-metaverse platforms. As an example, Sandbox offers a blockchain-based gaming, virtual plots and shopping. There is also an NFT (Non fungible token) world that belongs to its own world. It gives the users an opportunity to create their own digital world. Metahero project comes to the fore with its 3D scanning feature by which the users can create their avatar in 16K quality and convert it to NFT.

The Metaverse, which is introduced as a new perspective of the Internet, is also emerging as a good business opportunity in many sectors. Based on this, it is not difficult to predict how many financial opportunities it will turn into. Some of the leading technology companies believe that the metaverse has a future and is investing heavily in the technology. Examples of these companies include Meta, Microsoft, and Epic Games. There is a need for various technological developments in wearable devices, network connection technologies, etc. to realize the metaverse vision. Metaverse standards forum (https://metaverse-standards.org/) is being formed and there are studies on open standards. There have been advancements in virtual reality (VR) and augmented reality (AR) technologies lately. However, security and privacy concerns arise when data flow from sensory systems is used with various technologies and advanced algorithms. This paper aims to examine the possible security and privacy vulnerabilities of the metaverse and analyzes the solutions mainly in the authentication methods. 


In the next section, fundamentals of metaverse and virtual reality will be given. Privacy and security issues will be discussed in the third and fourth section. In the sixth section vr authentication methods  is given. Discussion is given in the seventh section. Conclusion and future works will be given in the last section.

\section{FUNDAMENTALS}

\subsection{Virtual Reality}
VR (Virtual Reality), in the simplest sense, is the creation of the world we see with our eyes in 3D with a computer. In order for the user to interact with this virtual world, a VR headset is used. Gloves or special clothing can be used to increase the reality and the immersiveness. The sensors can detect the user's movements, an illusion of "being there" (telepresence) is created simultaneously \cite{ref1}. In this way, it is possible for the user to move objects in the environment or to look at another environment with head movements. Virtual reality creates 3D spaces for us, while augmented reality is a bridge between this virtual world and the physical world.


\subsection{Metaverse}
Some see the Metaverse the improved version of virtual reality technology, as a fictional universe. However metaverse is an umbrella term for the future Internet which will consist of virtual worlds that are called verses. There are virtual worlds in computer games, but the specific characteristics of the metaverse make the difference. Metaverse should serve interesting experiences to attract a wide range of attendance. All the user operations should be synchronized and alive for the best user experience. The sustainablity of the system can be accomplished with the token economy \cite{ref10}. Decentralization will be needed to ensure the integrity of the system, token economy and decentralized identity. The digital assets will be kept in wallets and used in other verses with interoperability features. User patterns and digital assets will become more important to preserve and secure. We will need better security especially in the authentication phase in metaverse. 

\subsection{Security and Privacy of Authentication}
Authentication is the process of proving who a user or program is when accessing an environment. As the user gains access to all the assets after authentication; security and privacy must be assured. Security is the protection of the personal rights and human dignity of individuals in society, their property, from all kinds of dangers and accidents. Privacy is the name we give to the situation in which information, transaction data or correspondence belonging to the parties involved in a transaction is kept confidential from those outside the subject. Preserving the privacy of the personal data becomes harder as we become more connected with the technology. Body scanning, facial recognition software, DNA identification and retinal recognition are some of the techniques that can be used to ensure the user identity in metaverse \cite{ref2}. These are personal data, and ensuring the confidentiality and integrity of biometric data in this phase becomes more critical.

\section{PRIVACY ISSUES}

Millions of users share many of their data, including their private lives, on social media. These data reveal our political views, our family, our work, the things we love, in short, our lives. This situation was clearly demonstrated in several studies. Also the Web 2.0 technologies allowed the web developers to collect various user information and form user patterns. The forecasting ability derived from data is improving exponentially with the amount of data collected. Considering the amount of data that is collected by the social networking platforms, the amount of data that will be collected by metaverse will be much higher. The metaverse will be able to easily observe the movements of our body, any physiological response, and even brain waves. Preseving the confidentiality of these collected data becomes an important issue then. These data can be comprimised and the user privacy will be in risk.

The collected data in Metaverse can be summarized in three categories; personal information, behavior and communication patterns. Information from social networking platforms can be used to expose people's private lives, that is called doxing. There is a danger of accessing more sensitive information about the user such as the users' habits and physiological characteristics through the Metaverse.

Social engineering attacks account for the largest share of online cyberattacks, as measured during the COVID-19 pandemic\cite{ref3}. Social security attacks can become more powerful and easier with Metaverse. As the user will be mentally connected, the pyscological attacks may be more dangerous. Pyscological attacks can be carried with several methods such as espionage, stalking and alike. These can be avoided to some extent in the real world, but it may not be that easier to evade in the metaverse world. Implementing deterrent punishments will remain as a challenge in the metaverse world.

Considering the privacy risks in the Metaverse, several methods are being proposed to prevent these risks. The user is given an ability to make multiple clones of his avatar to disguise and relocate the user. These multiple clones can be created by teleporting an avatar and prevent unwanted tracking. However, companies and governments will still want to track the user. The conditions of this can be specified by the smart contracts.

\section{SECURITY ISSUES}

Many security threats and issues are present when several technologies are used i n the metaverse. Selected issues will be given in this section.

\subsection{Integrity and distinguishing a software agent:}
Metaverse integrity and authentication are one of the most important problems that exist. Data integrity ensures the protection and assurance of the accuracy and consistency of data throughout the entire life cycle. As an example, data integrity is critical in hospital information systems.
Data integrity becomes more critical in many cases and especially the risks inherent in machine-side transactions should be taken into consideration seriously. As an example, our fingerprint can be accessed even from photos we share on social media\cite{ref4}. 

Can we distinguish a software bot from a human? The Turing test was 
to test the intelligent behavior of a computer and see if we can distinguish it from a human. We will possibly interact with software guided by artificial intelligence in the metaverse. It is likely that we will be guided by having a bot with us in various activities (chatting, shopping). We will not be able to distinguish between artificial intelligence and humans at some point soon \cite{ref5}. Future attacks guided by artificial intelligence can be possible, and smart contracts can be used to control the systems.

\subsection{Human diversity in a single world:}
While the vulnerabilities so far depend on the platform's algorithms, the lack of a metaverse peer will also pose a security problem. There are many services in the metaverse. The web is very multiple, and people from all walks of life can find a suitable community of people for themselves. Everyone has the opportunity to turn to a platform to their liking. If we talk about the Metaverse, then people with many different thoughts and lifestyles will exist in the same place with each other at the same time. Although, they have acquired online communities for themselves, the diversity in this virtual environment cannot be as much as on the Web. People can use this platform to realize their terrible ambitions of bullying and many more. Many people from different types of life exist in the same environment, and bullying and even fraud are inevitable.

\subsection{VR Headset Security:}
The moment you plug in a VR headset and connect it to the internet, you will have a deep introduction to the digital world. When we think of virtual reality, we can actually think of it as the collection of biometric data. We are aware that fingerprint, voice, face recognition, retina scanning collect our data. Our sensitive data is stored somewhere. This sensitive information can be stolen, sold and used for criminal purposes. Similar avatar can be created in the metaverse by stealing our biometric sensitive data. This avatar can be used to commit human deception, crimes, or spread false information.

An attacker who has hacked your VR headset, will have the ability to see your office, surroundings, and even your bedroom from your camera. The attacker can also manage what you see, what you hear with the overlay attacks. Companies may promise privacy policies to protect their users from such attacks, but not applicable in many situations. The sensitive information (password, face recognition, fingerprint, etc.) that we describe in this section is actually the information that is required during authentication. It is not possible to enter the metaverse without these information.

\section{VR AUTHENTICATION METHODS}

Selected VR authentication methods in the literature are given in the following sub-sections. 

\subsection{Information-Based Authentication:}
This method is the most commonly used authentication method in the virtual reality environments. Verification is provided by entering a PIN or alphanumeric password before logging into the Metaverse universe. Various studies such as \cite{ref6} have been conducted to test this method. As it is shown in Figure 1 and Figure 2, 3D patterns, pattern lock, and PIN systems have been studied.

The test phase, which has two stages, analyzes both usefulness and safety. For convenience, the participants in the experiment are asked to memorize the password and enter it five times. The verification time and error rate are calculated by examining the input records. The purpose of the security test is to test what kind of privacy it has against shoulder surfing. The predictability of the password was measured by monitoring the hand movements of the person entering the password. In experiments conducted on these systems, it has been proven that 3D pattern is the safest. However it is not that the usability as it does not have such an easy use as a PIN system. Better interface designs must be made to correct the inverse ratio between usability and reliability.

\begin{figure}[ht]
\begin{center}
\includegraphics[width=7 cm]{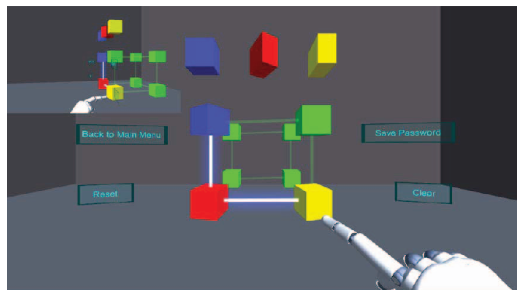}
\end{center}
\caption{User Interface of Authentication with 3D Patterns}
\label{figure1}
\end{figure}

\begin{figure}[h]
\begin{center}
\includegraphics[width=7 cm]{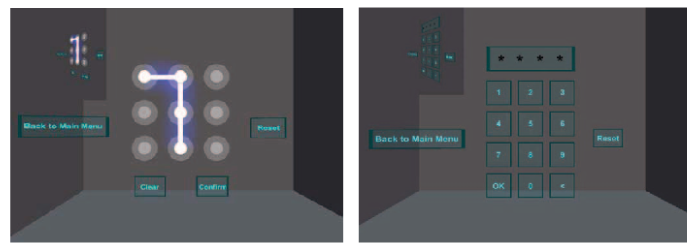}
\end{center}
\caption{User Interface of (1) Pattern Lock and (2) PIN System}
\label{figure2}
\end{figure}

\subsection{Biometric Authentication:}
Biometric data is used biometric authentication. There are many data types to be used in this type of verification, however Electroencephalography (EEG), body movements, and Electrooculography (EOG) readings are among the most used. EEG data is reliable because it is unique. In one study, subjects were shown a video both with and without VR. 8-channel EEG sensors and a Cyton board were used to receive data.

There is no significant difference in the results of the experiments. However, a model for the verification of brain signals is obtained in the metaverse, as also shown in Figure 3. This model had an accuracy of 80.91 \% \cite{ref7}. However, converting the biometric data of the users into data is enough to create a vulnerability.

\begin{figure}[ht]
\begin{center}
\includegraphics[width=5 cm]{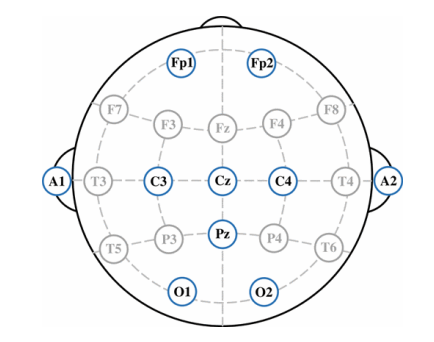}
\end{center}
\caption{Chosen Electrode Locations}
\label{figure3}
\end{figure}

\subsection{Multi-model Authentication:}
This method allows users to log in using a combination of two or more techniques during authentication. The security of the system is increased as it requires an attacker to bypass multiple security instead of one. The RubikBiom developed in this field can be an example of this model \cite{ref9}. Biometric behaviors collected from the user during authentication are controlled by the password entered in the rubik cube. As it is shown in Figure 4, the user selects the pin on the rubik's cube with the help of the vr control and the pattern of their biometric movements are controlled. Matching the password with the user's actions increases security. Multi-model systems are aimed at closing the vulnerabilities of a single authentication model.

\begin{figure}[ht]
\begin{center}
\includegraphics[width=5 cm]{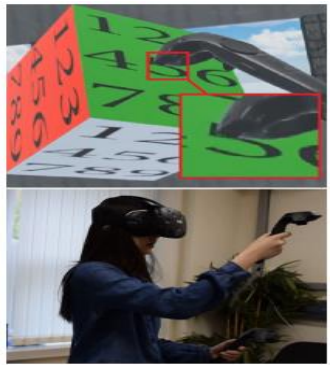}
\end{center}
\caption{Authentication with the pin on the rubik’s cube}
\label{figure4}
\end{figure}

Gaze-Based Authentication is a variant of multi-model authentication. In this method, authentication is performed using the human gaze. For example, if we want to scan a fingerprint, this needs to be done with a special device, or login can be done using any face image for face scanning. The eye movements are unique. It is possible to identify the user when examined together with extraocular muscle activations, Studies have been conducted that it would be useful to use ocular biomechanical analysis. During a video that is watched to the user, the eye sacs are monitored. Gaze control is performed by finding the eye joint angles (horizontal(H), vertical(V), torsional(T)). This can also be checked when the user enters a password with the VR headset \cite{ref8}. Since it is impossible to have any idea by watching from the outside, the possibility of imitation has disappeared with this method. It has been calculated that the error rate is very low and the average input time is 5.94 seconds \cite{ref11}.

\section{DISCUSSION}

The advantages and disadvantages of the investigated methods vary according to various parameters. The investigated methods are compared in Table 1.

\begin{table}[htbp]
\caption{Advantages and Disadvantages of Authentication Mechanisms}
\begin{center}
\begin{tabular} {| p{0.25\linewidth} | p{0.25\linewidth} | p{0.25\linewidth} |}
\hline
\textbf{\textit{METHODS}}& \textbf{\textit{ADVANTAGES}}& \textbf{\textit{DISADVANTAGES}} \\
\hline
Information-Based\\ Authentication:& ease of use & low reliability \\
\hline
Biometric\\Authentication:&ease of authentication& low reliability \\
\hline
Multi-model\\ Authentication:&secure& disclosure threat of biometric data \\
\hline
Gaze-Based\\ Authentication:&unique& disclosure threat of biological data \\
\hline
\end{tabular}
\label{tab1}
\end{center}
\end{table}

In the information-based authentication method, the user is alone with a system that he will use very easily. The quick completion of the process is also a significant plus of this method. However, since no personal data is entered, the password to be entered by the user can be easily obtained by methods such as shoulder-surfing. Since the user cannot see the physical world due to the VR headset, care should be taken in this regard.

In biometric authentication, there is no possibility of imitating authentication data. In this context, it is certainly more reliable than the first method we mentioned. We cannot say that the model created using our biometric data is completely accurate either. The disadvantage of this method is that extracting our brain model and stored it elsewhere. If this data is not protected with cryptographic methods, the compromise of this data can create privacy problems.

In the multi-modal method, multiple verification models are used to enter into the virtual world. The user performs a unique verification with gaze-based authentication in a way that can not be understood by the people in the same physical space. The highest reliability is ensured when this method is combined with a predetermined schematic image.

The multi-modal authentication is by far the most reliable among the methods we investigated. But work on behalf of people with various physical disabilities and the elderly should also be diversified. For example, how to tolerate visual impairments? What are the effects of visual impairment during authentication? There is a need for more studies on these questions. For the elderly, on the other hand, various limbs are difficult to control, so studies can be carried out on this, eliminating factors that threaten the safety of authentication.

\section*{CONCLUSION}

The metaverse is making a rapid entry into our lives with the current technology developments. 
People's desire to share experiences and the life of society will increase with this virtual reality worlds. Privacy and security threats are examined and authentication mechanisms are compared in this study. Biometric authentication is strong but the biometric data has to be kept secure. Multi-model authentication seems the most reliable method of all. 

Authentication does not seem to have been sufficiently overemphasized in the context of VR. The issues of how the authentication process can best be integrated into VR environments and how it can work for users while wearing VR glasses should be taken into account. The combination of eye-gaze knowledge and information-based authentication in this area promises a future.

There is a need for more detailed studies on this concept. Blockchain and smart contract based decentralized techniques can be used to enhance the integrity of the data and ensure that artificial techniques are being used as it should.

\end{document}